# PILOT TONE-GUIDED FOCUSED NAVIGATION FOR FREE-BREATHING WHOLE-LIVER FAT-WATER AND T2* QUANTIFICATION

Adèle LC Mackowiak[1,2,3], Christopher W Roy[1], Mariana BL Falcão[1], Mario Bacher[1,6], Aurélien Bustin[1,4,5], Jérôme Yerly[1,7], Peter Speier[6], Matthias Stuber[1,7], Naïk Vietti-Violi[1], Jessica AM Bastiaansen[2,3]

[1]Department of Radiology, Lausanne University Hospital (CHUV) and University of Lausanne (UNIL), Lausanne, Switzerland

[2]Department of Diagnostic, Interventional and Pediatric Radiology (DIPR), Inselspital, Bern University Hospital, University of Bern, Switzerland

[3]Translational Imaging Center (TIC), Swiss Institute for Translational and Entrepreneurial Medicine, Bern, Switzerland

[4]Department of Cardiovascular Imaging, Hôpital Cardiologique du Haut-Lévêque, CHU de Bordeaux, France

[5]IHU LIRYC, Electrophysiology and Heart Modelling Institute, INSERM U1045, Centre de Recherche Cardio-Thoracique de Bordeaux, France

[6]Siemens Healthcare GmbH, Erlangen, Germany

[7]CIBM Center for Biomedical Imaging, Lausanne, Switzerland

**Corresponding author**: Jessica A.M. Bastiaansen, Ph.D., Assist. Prof.

Email: jbastiaansen.mri@gmail.com

Freiburgstrasse 3, 3010 Bern, Switzerland. Department of Diagnostic, Interventional and Pediatric Radiology (DIPR), Inselspital, Bern University Hospital, University of Bern, Switzerland

# ABSTRACT

## *Purpose*

To achieve whole-liver motion-corrected fat fraction (FF) and R2* quantification with a 3-minute free-breathing (FB) 3D radial isotropic acquisition, for increased organ coverage, ease-of-use, and patient comfort.

## *Methods*

A FB 3D radial multiecho gradient-echo liver acquisition with integrated Pilot Tone (PT) navigation and $N_{TE}$=8 echoes was reconstructed with a motion-correction algorithm based on focused navigation and guided by PT signals (PT-fNAV), with and without a denoising step. Fat fraction (FF) and R2* quantification using a graph cut algorithm was performed on the motion-corrected whole-liver multiecho volumes. Volunteer experiments (n=10) at 1.5T included reference 3D and 2D Cartesian breath-hold (BH) acquisitions. Image sharpness was assessed to evaluate the quality of motion correction with PT-fNAV, compared to a motion-resolved reconstruction. Fat-water images and parametric maps were compared to BH reference acquisitions following Cartesian trajectories, and to a routinely used clinical software (MRQuantiF).

## *Results*

The image sharpness provided by PT-fNAV (with and without denoising) was similar in end-expiratory motion-resolved reconstructions. The 3D radial FB FF maps compared well with reference BH 3D Cartesian maps (bias +0.7%, limits of agreement (LOA) [-2.5; 4.0]%) and with 2D quantification with MRQuantiF (-0.2%, LOA [-1.1; 0.6]%). While expected visual deviations between proposed FB and reference BH R2* maps were observed, no significant differences were found in quantitative analyses.

## *Conclusion*

A 3D radial technique with retrospective motion correction by PT-fNAV enabled FF and R2* quantification of the whole-liver at 1.5T. The FB whole-liver acquisition at isotropic spatial resolution compared in accuracy with BH techniques, enabling 3D assessment of steatosis in individuals with limited respiratory capabilities.

## INTRODUCTION

The incidence of fatty liver diseases is rapidly increasing worldwide[1], particularly in the context of the obesity epidemic[2]. Chemical shift-encoded MRI (CSE-MRI) enables the noninvasive quantification of fat[3–5], providing an accurate alternative to gold-standard liver biopsies[6–9], which is crucial for monitoring the treatment of hepatic steatosis[10,11]. CSE-MRI is typically performed with multiecho gradient-echo (ME-GRE) sequences, requiring the acquisition of a large number of echoes to resolve the fat spectrum in its entirety[3,12] and to quantify R2*[13], which is a metric for excessive iron deposition[14]. Unlike focal biopsies, CSE-MRI is not limited in organ coverage and can account for steatosis heterogeneity[15,16].

Traditional CSE-MRI uses Cartesian sampling and breath-hold (BH) techniques, which are sensitive to physiological motion and can be challenging due to patient compliance. The volumetric interpolated BH examination (VIBE) technique[17–19] necessitates multiple slices for whole-organ coverage, leading to repeated BH or long scans. Free-breathing (FB) techniques increase ease-of-use, scanning efficiency, and patient comfort, making them particularly beneficial for populations where BHs are not sustainable. Additionally, FB techniques enable accurate quantitative imaging[20–23], which often requires longer scan times.

In contrast to Cartesian sampling, CSE-MRI using radial k-space trajectories is inherently robust to motion, which can often be extracted from the data itself[24]. Advanced motion-resolved approaches using compressed sensing[25–27] (CS), or self-gating (SG) approaches in multiecho GRE acquisitions[21,28–31], have surpassed the motion averaging methods initially employed[27,32,33]. Nevertheless, the parametric maps obtained with the aforementioned studies focused on 1-dimensional translational motion of the liver, potentially neglecting the more complex respiratory motion patterns of the organ.

Recently, a focused navigation (fNAV) algorithm was introduced[34], based on autofocusing[35,36], which extracts 3D displacement fields of both rigid and non-rigid motion. fNAV abbreviated reconstruction time in comparison to CS techniques when applied to the heart[34,37]. In parallel, 3D radial golden-angle trajectories facilitated highly under-sampled FB ME-GRE acquisitions, allowing for isotropic fat-water quantification in large volumes[38]. In this prior study[38], improved motion correction was achieved when combined with Pilot Tone[39] (PT) navigation, which can benefit low SNR sequences like GRE Therefore, it is hypothesized that by leveraging a 3D radial trajectory, a non-rigid motion-correction algorithm, and PT navigation, whole-liver quantification of FF and R2* of similar accuracy to BH reference scans can be obtained under free-breathing conditions.

The goal of the study was to address the need for whole-liver free-breathing quantitative imaging, by applying PT-guided fNAV reconstruction to 3D radial FB ME-GRE acquisitions. Whole-liver FF and R2* quantification was performed in volunteers at 1.5T. The impact of motion correction with PT-fNAV on parametric mapping, and the agreement of 3D radial FB motion-corrected images and maps with reference BH Cartesian acquisitions, were determined. A comparison with FF and R2* estimates from a routinely used MRQuantiF clinical software for iron and fat quantification was provided.

# METHODS

## *2.1. MRI experiments*

Ten healthy subjects enrolled in the study (F=6, median age [± interquartile range]: 26.5 [±3.25] y.o., body mass index: 22.7 [±3.4]) after consent was collected following institutional rules (authorization CER-VD 2021-007008 Switzerland). Experiments were performed on a 1.5T clinical scanner (MAGNETOM Sola, Siemens Healthcare, Erlangen, Germany) using a 12-channel body array with a PT generator (Biomatrix Body 12, Siemens Healthcare, Erlangen, Germany). A free-breathing 3D radial ME-GRE sequence[38] with $N_{TE}$=8 echoes was used. The spiral phyllotaxis[40] trajectory used consisted of 424 interleaved spirals of 22 readout lines each, resulting in a dataset under-sampled to 16% of the Nyquist criterion, corresponding to an acceleration factor R=6.21 and leading to a fixed scan time TA=2:57min in all subjects. Reference acquisitions included a 3D ME-GRE scan with 8 slices (3D Cartesian BH), a 2D 2-point Dixon VIBE scan (2D Cartesian 2p-Dixon VIBE BH), and a 2D ME-GRE scan with $N_{TE}$=10 echoes (2D MRQuantiF BH, recommended protocol for processing with MRQuantiF software, see https://imagemed.univ-rennes1.fr/en/mrquantif/overview), all acquired with a 5mm slice thickness in a single BH. Sequence parameters are detailed in **Table 1**.

All the BH protocols were performed either at end-expiration (n=6/10) or end-inspiration according to the subject's capability. Images from the reference breath-held scans (3D Cartesian BH, 2D 2p-Dixon VIBE BH, and 2D MRQuantiF BH) were reconstructed at the scanner using built-in vendor software.

| Sequence parameters | 3D radial FB | 3D Cartesian BH | 2D Cartesian 2p-Dixon VIBE BH | 2D MRQuantiF BH |
|---|---|---|---|---|
| **Respiratory condition** | Free breathing | Breath hold | Breath hold | Breath hold |
| **FOV size [mm²]** | 290*290 | 290*290 | 290*290 | 290*290 |
| **FOV phase [%]** | 100 | 87.5 | 81.3 | 87.5 |
| **Spatial resolution [mm³]** | 1.5*1.5*1.5 | 1.5*1.5*5.0 | 1.5*1.5*5.0 | 1.5*1.5*5.0 |
| **Number of slices** | 192 | 8 | 8 | 3 |
| **Slab coverage [cm]** | 28.8 | 4 | 4 | 1.5 |
| **Readout trajectory** | phyllotaxis[40] | Cartesian | Cartesian | Cartesian |
| **TR [ms]** | 19.02 | 19.24 | 6.70 | 120.00 |
| **$\Delta TE/TE_1$ [ms]** | 2.30/1.38 | 2.30/1.30 | 2.38/2.39 | 2.38/2.38 |
| **Number of echoes $N_{TE}$** | 8 | 8 | 2 | 10 |
| **Flip angle [°]** | 12 | 12 | 10 | 20 |
| **Pixel bandwidth [Hz/px]** | 898 | 898 | 470 | 530 |
| **Gradient readout mode** | monopolar | monopolar | bipolar | bipolar |
| **Acceleration factor R** | 6.2 (fNAV) 24.8 (motion-resolved) | none | 4 | none |
| **TA [min]** | 02:57 | 00:20 | 00:07 | 00:15 |

**Table 1** **Sequence parameters for MRI volunteer experiments at 1.5T.**

Abbreviations: BH, breath-hold; fNAV, focused navigation; FOV, field-of-view; FB, free-breathing; TA, acquisition time; TE, echo time; TR, repetition time.

## *2.2. 3D radial free-breathing image reconstruction*

All reconstruction steps were performed in MATLAB R2021b (The Mathworks, Natick, MA, USA). A 1D respiratory signal trace was extracted[41] from the PT signals sampled at a 2kHz rate throughout the acquisition (**Figure 1**). The end-expiratory position in PT signals was identified and chosen to guide all fNAV reconstructions. Respiration was also measured by a respiratory sensor (RSR) built into the scanner, as a reference. The fNAV iterative reconstruction used a non-uniform FFT and coil sensitivity profiles. At each iteration, 3D amplitude coefficients $\mathbf{A}(\mathbf{r})$ are computed at each location $\mathbf{r}$ and updated following a metric of gradient entropy[35] for image blur assessment. The initial coefficients were restricted to rigid displacements in a selected ROI, and non-rigid motion was modeled afterwards as a sum of rigid displacements. The final corrected volumes are obtained when both the entropy metric converges to a minimum and a smoothness constraint applied on the displacement field is optimized.

The correction coefficients determined with fNAV were applied to each set of echoes, and 3D coil sensitivities were computed using data from the first echo. Following fNAV reconstruction, the motion-corrected multiecho volumes were denoised using multi-contrast patch-based denoising[42], with a regularization coefficient of σ=75 and a patch size of 9x9 pixels. Computation times were recorded.

To evaluate image quality concerning respiratory motion blurring, sharpness measurements at the lung-liver interface were performed on the 3D radial FB images, using the slope of parametrized sigmoid functions[43]. The analysis involved comparing images reconstructed under four methods: 1) without any motion compensation, 2) with PT-fNAV, 3) with PT-fNAV + denoising, and 4) with a motion-resolved reconstruction based on CS, guided by the same PT respiratory signal but without denoising. For the fourth method, end-expiratory slices were selected that were obtained after data binning into 4 respiratory states, as previously described[44]. For each method, the same set of 20 coronal slices was analyzed. For each slice, sigmoid functions were fitted to 15 image segments perpendicular to the interface delineated with a manual selection and Bézier interpolation, and the mean sharpness per slice was reported. For all four methods, the mean sharpness and standard deviation across all slices and all volunteers were reported and statistical differences were assessed using paired Student's t-tests corrected for multiple comparisons, with $p<0.05$ considered significant.

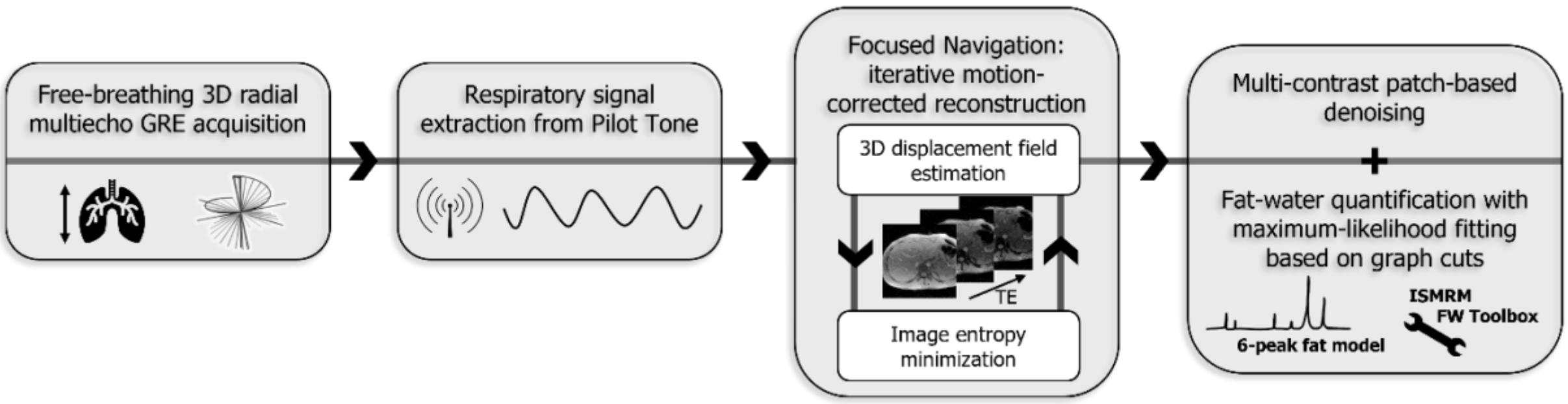

**Figure 1 Acquisition, reconstruction, and post-processing steps for 3D free-breathing motion-corrected whole-liver fat-water quantification**

Whole-liver data is collected during free-breathing with a 3D multiecho GRE sequence following a spiral phyllotaxis readout trajectory. The coil-integrated Pilot Tone navigator allows the retrospective monitoring of the chest position via the extraction of a 1D respiratory signal trace. This extracted respiratory signal is used to guide fNAV for motion-corrected image reconstruction. Then, multiecho imaging volumes are post-processed with patch-based denoising, and FF and R2* maps are generated using maximum-likelihood fitting and a reference 6-peak fat model.

## *2.3. Fat fraction and R2* quantification and analysis*

The denoised PT-fNAV multiecho volumes were selected for fat fraction and R2* quantification. A maximum-likelihood algorithm based on graph cuts[45,46] was used to quantify FF, $B_0$ and R2* and to obtain fat-only and water-only images. Fitting parameters included: a range of [-350;350] Hz for $B_0$ map estimation, a range of [0;100] Hz for R2* estimation with a single T2* decay component, n=50 graph cuts iterations, and a regularization term λ=0.05. A 6-peak liver fat spectral model[47] was used as input. An automatically generated background mask was placed on the denoised volumes, to accelerate the computation of the separated images and maps with the fitting algorithm. To provide reference fat-water images and maps, the 3D Cartesian BH multiecho images were reconstructed at the scanner and then post-processed using the same parameters. To account for the relatively large RF excitation angle (α=12°) a T1 bias correction[48] was performed on the FF maps (**Supporting Information S1**) following the methodology of Yang et al.[49]

The fat-only and water-only images were qualitatively compared in terms of apparent species separation and contrast to the corresponding images obtained with the 3D Cartesian BH and the 2D 2-point Dixon VIBE BH scans. For quantitative comparisons, the FF and R2* maps produced from 3D radial FB and 3D Cartesian BH scans were assessed via ROI-based analyses. Note that with only 2 echoes, Dixon VIBE protocol cannot produce reliable parametric maps, therefore was not included in the comparison. An experienced abdominal radiologist (NVV) drew ROIs in three liver segments on the 2D MRQuantiF BH images. From these, the MRQuantiF software provided one mean FF and one mean R2* value per ROI (but no full parametric map, see **Supporting Information S2**). Then, the same ROIs were used to analyze the maps obtained from the 3D datasets. Maps exhibiting large and/or abrupt deviations towards extreme values (see **Supporting Information S3**) were reported as corrupted and discarded from the analysis when they occurred. Bland-Altman analyses and paired Student's t-test were performed to the precision and bias of the proposed 3D radial FB maps compared to the 3D Cartesian BH protocol and the MRQuantiF measurements.

# RESULTS

## *3.1. Validation of PT-fNAV for respiratory motion correction of 3D radial ME-GRE free-breathing data*

In all volunteers, the respiratory signal curves extracted from the PT signals compared well with amplitude variations recorded by the respiration sensor (**Figure 2B**). The average image reconstruction time was 37min (PT-fNAV, with a range of [14-62]min), 48min (with added patch-based denoising, with a range of [15-73]min), 6.4h (motion-resolved, with a range of [4.9-14.5]h). PT-fNAV, both with and without denoising, improved image quality with minimized motion-induced blurring at the lung-liver interface (**Figure 2A**, yellow arrows) and at the location of hepatic veins (**Figure 2A**, white arrows) in comparison to un-corrected images. Motion-resolved images, without denoising, appeared noisier across the whole FOV than PT-fNAV reconstructions without denoising (**Figure 2A**). Notably, the hepatic veins were more visible with PT-fNAV without denoising than with the CS-based reconstruction (Figure 2A, white arrows). Across the volunteer cohort, the average and standard deviation of the sharpness at the lung-liver interface was (0.155±0.105) without motion compensation, (0.693±0.104) with PT-fNAV, (0.741±0.149) with PT-fNAV + denoising and (0.821±0.175) with the motion-resolved reconstruction. While all three reconstruction methods showed a significant improvement in lung-liver interface sharpness ($p<0.01$) compared to the non-compensated reconstruction, no significant differences in sharpness were found between any of the three techniques (PT-fNAV vs PT-fNAV + denoising: $p=0.41$, PT-fNAV vs PT motion-resolved: $p=0.36$, PT-fNAV + denoising vs PT motion-resolved: $p=0.53$). While the end-expiratory motion-resolved images scored higher in sharpness, they displayed higher intra- and inter-volunteer variability (**Figure 2C**). It is to be noted that inter-volunteer variability followed a similar order across the techniques (e.g., in **Figure 2C**, V6 consistently scored the lowest).

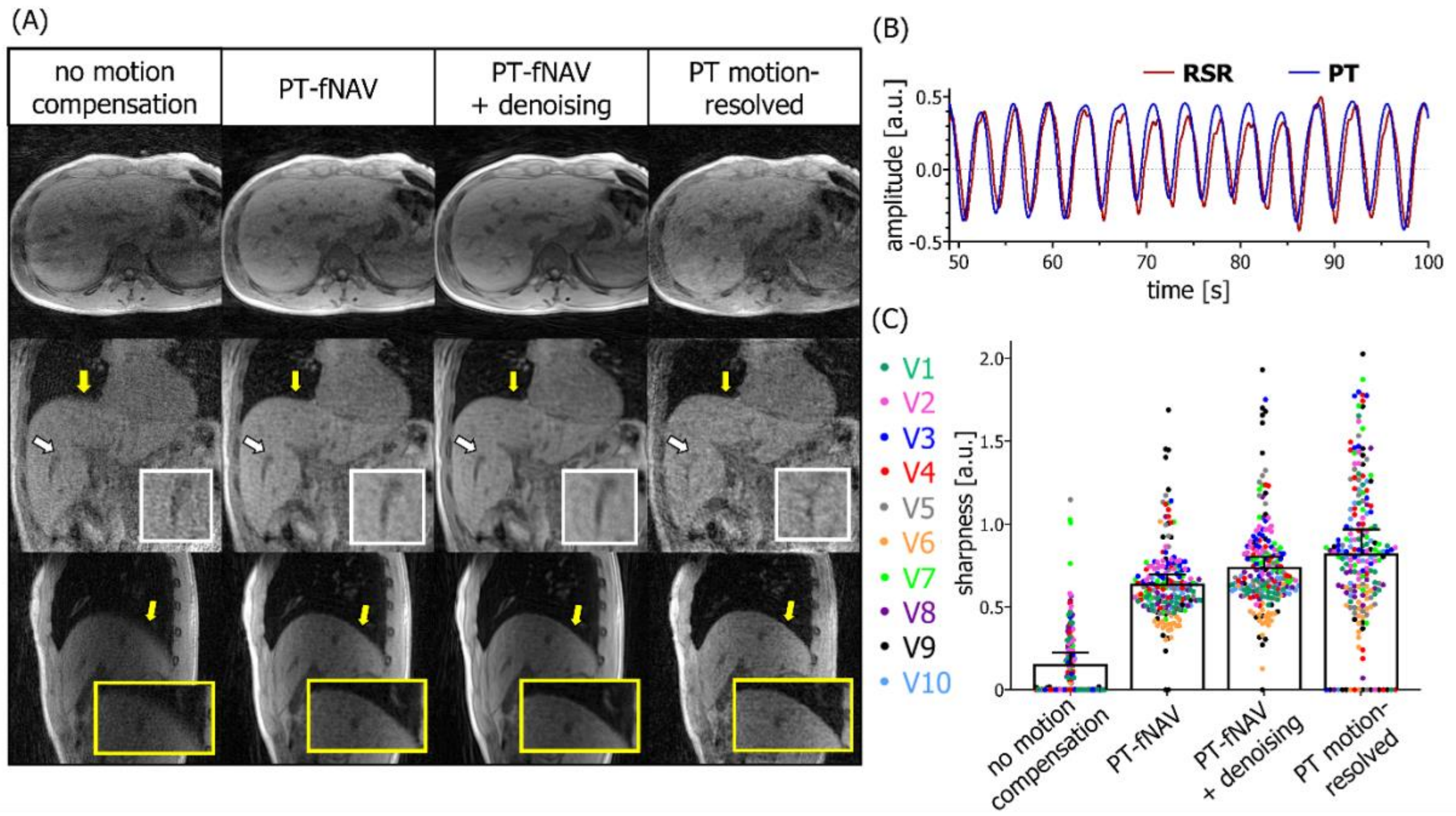

**Figure 2 Different types of PT-driven respiratory motion compensation on 3D radial free-breathing multiecho data**

(A) Image reconstructions of the 3D radial data collected at the first TE without motion compensation (simple gridding procedure with coil sensitivity estimation), with the PT- fNAV reconstruction (with and without denoising) and the motion-resolved CS reconstruction are shown in one volunteer. The resolution of details such as hepatic veins (white arrows) was improved with PT-fNAV, even without denoising, compared to the CS reconstruction (B) The corresponding Pilot Tone respiratory signal trace is shown next to the amplitude deviations measured by the respiratory sensor (RSR). (C) Sharpness measurements at the lung-liver interface for the 4 reconstruction techniques are shown for all n=10 volunteers (one color per volunteer, one dot per slice). Bars indicate the mean and standard deviation of the sharpness obtained in each reconstruction technique.

## *3.2. Fat-water separation and parametric quantification*

In this section, we consider the fat-water separation and parametric quantification results obtained using the PT-fNAV + denoising reconstruction, which markedly improved image quality by effectively reducing motion artifacts in the water-fat images, the FF maps, and the R2* maps. No fat-water swaps were observed indicating a robust separation of chemical species (**Figure 3**). The water-only images obtained with the 3D radial FB scans displayed higher contrast between the hepatic vessels and the liver tissue than the images from 3D

Cartesian BH scans, thus comparing better to the 2D Cartesian VIBE images, which are characterized by lower fluid signal intensities in the hepatic veins.

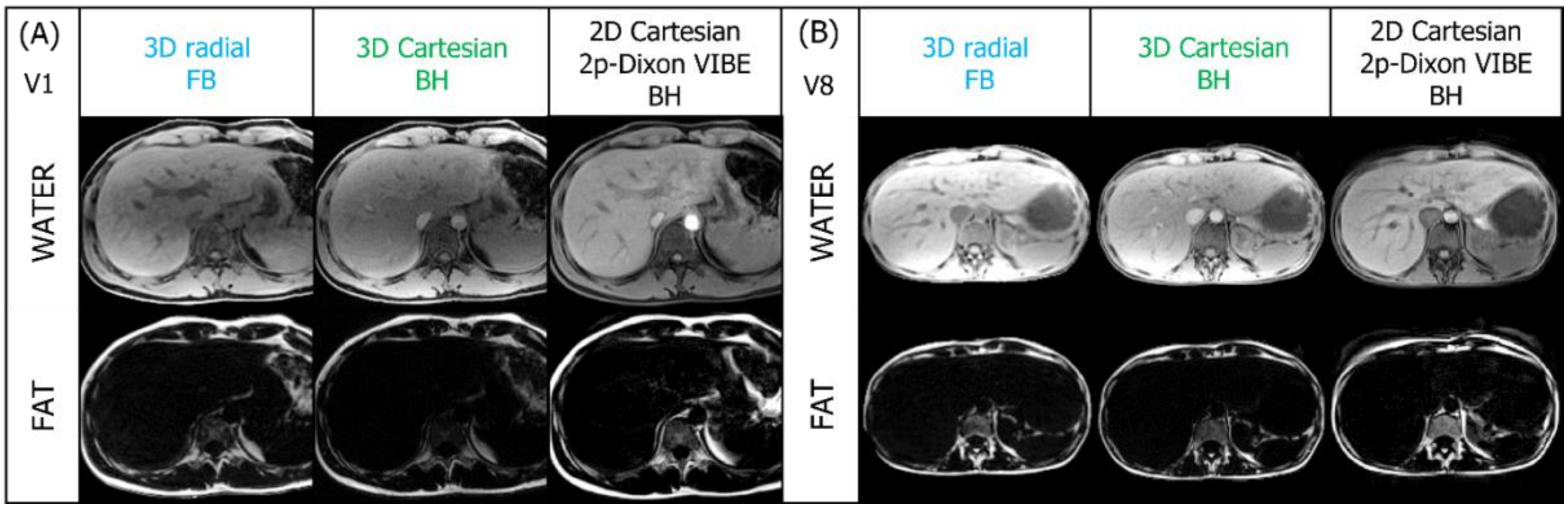

**Figure 3 Fat-water separation with 3D radial FB, 3D Cartesian BH, and 2D Cartesian BH sequences in 2 volunteers**

The water-only and fat-only images obtained with the 3D radial multiecho free-breathing sequence are compared to a 3D Cartesian breath-hold sequence with similar acquisition parameters. The first two columns of panels (A) and (B) (showing volunteers V1 and V8, respectively) show separated images obtained by fitting the reconstructed 8-echo datasets with a maximum-likelihood fitting routine based on graph cuts, using a spectral model of fat with 6 peaks. A second comparison is provided in the third column of each panel, in which water-fat separated images obtained with a routine 2-point 2D Cartesian VIBE BH sequence are displayed.

Throughout the volunteer cohort, out of the 80 slices collected with the 3D Cartesian BH scan, 4 FF maps and 15 R2* maps were reported as corrupted and discarded from quantitative analyses. It is to be noted that among the corrupted R2* maps, 60% occurred on a slice at the extremity of the 3D slab. No corrupted maps were reported within the 3D radial FB maps (see also **Supporting Information S3**).

In all other slices, FF maps from 3D radial data visually compared well with their 3D Cartesian counterparts (**Figure 4A**). The FF values estimated outside of the liver parenchyma were within the expected ranges for healthy adults (e.g., FF of 90-100% in the sub-cutaneous fat) for all sets of 3D maps. Hepatic ROI-based analyses showed a bias of +0.7% in fat fraction with limits of agreements (LOA) at [-2.5; 4.0]% between the proposed 3D radial FB maps and the 3D Cartesian BH maps (**Figure 4C**), and the paired Student's t-test revealed a significant difference ($p<0.01$) between the two measurement techniques. Differences were mostly driven by one volunteer for which higher (>5%) fractions with high variability between slices and ROIs were detected (**Figure 4C**, V4). The comparison to FF measurements provided by the MRQuantiF software showed a bias of -0.9% with LOA [-1.7; 0.1]% for the 3D Cartesian maps and a bias of -0.2% with LOA [-1.1; 0.6]% for the 3D radial maps (**Figure 4E**). Detailed FF measurements are provided in table format in **Supporting Information S4**.

While the R2* values estimated with both 3D protocols ranged within the normal values at 1.5T[50], the global appearance of the maps differed more substantially (**Figure 4B**). Maps from the 3D radial FB scans showed overall larger distributions in R2* across the different liver segments, with in some cases high values estimated at the posterior edge of the lobe, due to magnetic susceptibility effects close to the air-filled lungs (**Figure 4B**, arrow). In comparison, R2* maps from Cartesian scans displayed more homogeneous values across the organ. Bland-Altman analyses showed a bias of $3.2s^{-1}$ with LOA [-12;19]% between R2* estimated from 3D Cartesian BH and 3D radial FB scans (**Figure 4D**). With respect to MRQuantiF values (**Figure 4F**), the R2* bias from 3D Cartesian BH scans was $0.2s^{-1}$ with LOA $[17; 17]s^{-1}$. 3D radial R2* maps showed a larger bias with MRQuantiF at $0.6s^{-1}$, but with tighter limits of agreement at $[-7.9; 6.7]s^{-1}$. The FF values obtained with the graphcut algorithm in the evaluated ROIs were statistically similar, both when comparing the 3D radial FB and the 3D Cartesian BH to MRQuantiF values and between themselves.

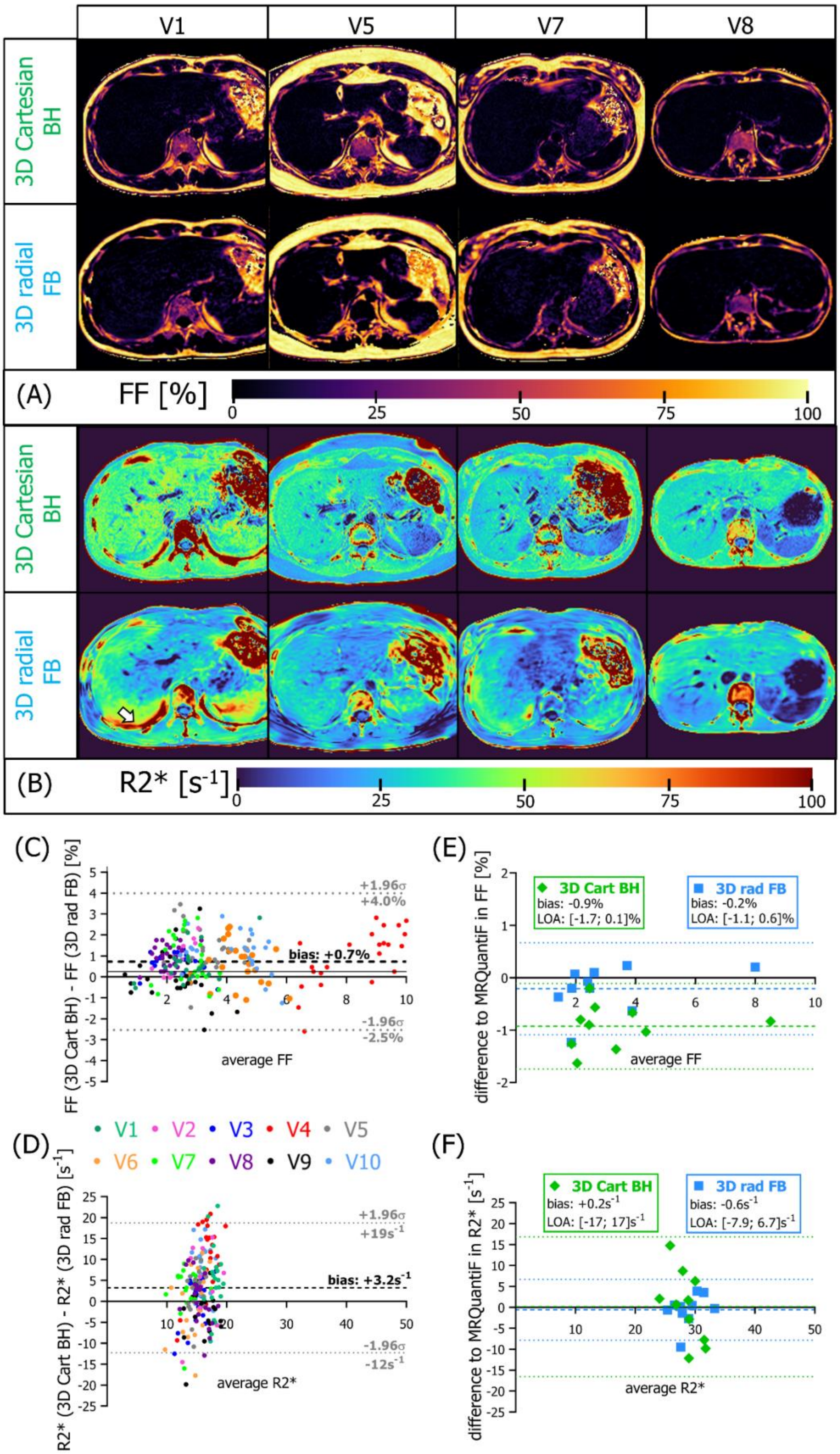
V1
V5
V7
V8
3D Cartesian BH
3D radial FB
(A)
FF [%]
0
25
50
75
100
3D Cartesian BH
3D radial FB
(B)
R2* [s-1]
0
25
50
75
100
(C)
FF (3D Cart BH) - FF (3D rad FB) [%]
+1.96σ
+4.0%
bias: +0.7%
-1.96σ
-2.5%
average FF
(E)
difference to MRQuantiF in FF [%]
3D Cart BH
bias: -0.9%
LOA: [-1.7; 0.1]%
3D rad FB
bias: -0.2%
LOA: [-1.1; 0.6]%
average FF
V1 V2 V3 V4 V5
V6 V7 V8 V9 V10
(D)
R2* (3D Cart BH) - R2* (3D rad FB) [s-1]
+1.96σ
+19s-1
bias: +3.2s-1
-1.96σ
-12s-1
average R2*
(F)
difference to MRQuantiF in R2* [s-1]
3D Cart BH
bias: +0.2s-1
LOA: [-17; 17]s-1
3D rad FB
bias: -0.6s-1
LOA: [-7.9; 6.7]s-1
average R2*

**Figure 4 Fat and R2* parametric maps and Bland-Altman analyses in volunteers**

Transversal fat fraction (A) and R2* (B) maps obtained with the graph cut algorithm from the proposed 3D radial free-breathing acquisition and the reference 3D Cartesian breath-hold scan in 4 volunteers (from left to right: V1, V5, V7, V8). The bias and limits of agreement (LOA) of the proposed radial technique with respect to the 3D Cartesian are reported for FF (C) and R2* (D). In (C) and (D), mean and standard deviation measurements were performed in 3 ROIs over a central slab (8 Cartesian slices, 20 radial slices) for each volunteer as indicated by the colour code. In (E) and (F), the bias and LOA of both 3D techniques with respect to the 2D MRQuantiF BH measurements are reported for FF and R2* for all 10 volunteers. The mean FF (E) and R2* (F) values measured over all ROIs and slices in the 3D maps are compared to the mean values provided by the MRQuantiF software.

# DISCUSSION

With the proposed approach, data to construct whole liver maps of FF and R2* was acquired in a scan time under 3min, thus clinically feasible[51], while providing additional echo sampling and a 3D isotropic resolution with a considerable improvement in ease-of-use for the examiner. A comprehensive comparison of four different image reconstruction techniques for whole-liver free-breathing imaging data, and a comparison with reference breath hold scans, revealed that the motion-corrected PT-fNAV reconstruction allowed for noise reduction and preserved edge sharpness at the lung-liver interface, even without performing additional denoising and compared with a compressed sensing-based motion-resolved approach. To our knowledge, this is the first use of fNAV and Pilot Tone for quantitative and structural whole-liver MRI.

Fat quantification using the proposed 3D radial approach agreed with values from reference scans. The calculated biases in FF quantification were in general below 1%, which is lower than the typically observed sensitivity range of CSE-MRI. These findings show that the proposed technique provides coherent 3D mapping volumes, with no mismatches between adjacent slices that can be reported with methods that use slice-encoding. The effect of susceptibility on R2* quantification[52] was observed both in end-inspiration BH Cartesian acquisitions with large air-filled regions and more so in 3D radial FB, at locations of expected higher motion amplitudes. This is in line with prior investigations which reported respiratory-induced R2* aliasing[11,28,29], and suggests potential residual motion artifacts and a higher sensitivity to motion compared to FF maps.

This work addresses the need for whole-liver fat quantification during free-breathing, thus enabling 3D monitoring of potentially inhomogeneous steatosis, even in patients with limited pulmonary capacities. Additionally, the integration of the sequence into the clinical practice is simplified due to the single-click approach of the 3D FB scan. The use of Pilot Tone technology to monitor liver motion may benefit low SNR 3D radial sequences instead of traditional self-gating[21,30] or similarity-driven binning algorithms[53]. It opens new avenues in trajectory design that more uniformly covers k-space and could improve robustness to gradient imperfections[54]. In addition, PT technology combined with free-breathing acquisitions will enable the translation of fat quantification methods that require a magnetization steady-state[55].

Still, the requirement for clinical adaptation of free-breathing quantification techniques is a rapid online image reconstruction. By opting for motion correction with fNAV and image-based parametric mapping, the computational burden is heavily reduced compared to CS-based reconstructions. Additionally, the choice of monopolar readouts avoid the need for gradient corrections[56–58] in the reconstruction pipeline. Furthermore, the denoising step did not significantly impact the overall results and could therefore be omitted from the reconstruction. With the emergence of open-source workflows[59,60] for image reconstruction, our proposed approach could be implemented on an MR scanner.

A comprehensive validation in fatty livers of available FB techniques, including comparisons between 3D radial and stack-of-stars trajectories that are more common for abdominal imaging[21,28,31–33], should be performed in future work as to complete the validation[63]. Nevertheless, this study performed in healthy subjects at 1.5T demonstrates the feasibility of obtaining accurate parametric maps of hepatic fat from FB scans with relatively accessible tools such as Pilot Tone for retrospective motion monitoring and an open-source fitting algorithm[46].

## CONCLUSION

In conclusion, this study validated a 3D radial FB pipeline with PT-fNAV motion correction for accurate fat quantification of the whole liver in under 3 minutes of scan time, with a user-friendly sequence. This approach presents a promising alternative to conventional 2D and 3D BH scans for populations with reduced respiratory capabilities.

## ACKNOWLEDGMENTS

We acknowledge the use of the ISMRM Fat–Water Toolbox (http://ismrm.org/workshops/FatWater12/data.htm) for some of the results shown in this article.

This study was supported by funding received from the Swiss National Science Foundation (grants PCEFP2_194296 and PZ00P3_167871 to J.A.M.B., grant PZ00P3_202140 to C.W.R., grants 320030B_201292 and 320030_173129 to M.S.), the University of Lausanne Bourse Pro-Femmes (J.A.M.B.), the Emma Muschamp Foundation (J.A.M.B.) and the Swiss Heart Foundation (grant FF18054 to J.A.M.B.).

## CONFLICT OF INTEREST STATEMENT

Mario Bacher and Peter Speier are employees of Siemens Healthcare AG. Matthias Stuber receives non-monetary research support form Siemens Healthineers.

## DATA AVAILABILITY STATEMENT

All anonymized volunteer datasets (n=10), consisting of (1) the 3D radial FB raw data files, (2) the raw Pilot Tone signals and the extracted respiratory signals, and (3) the reference images obtained from the 3D Cartesian BH and 2D 2p-Dixon VIBE sequences are available publicly at the following repository: https://zenodo.org/records/8338128.

# SUPPORTING INFORMATION

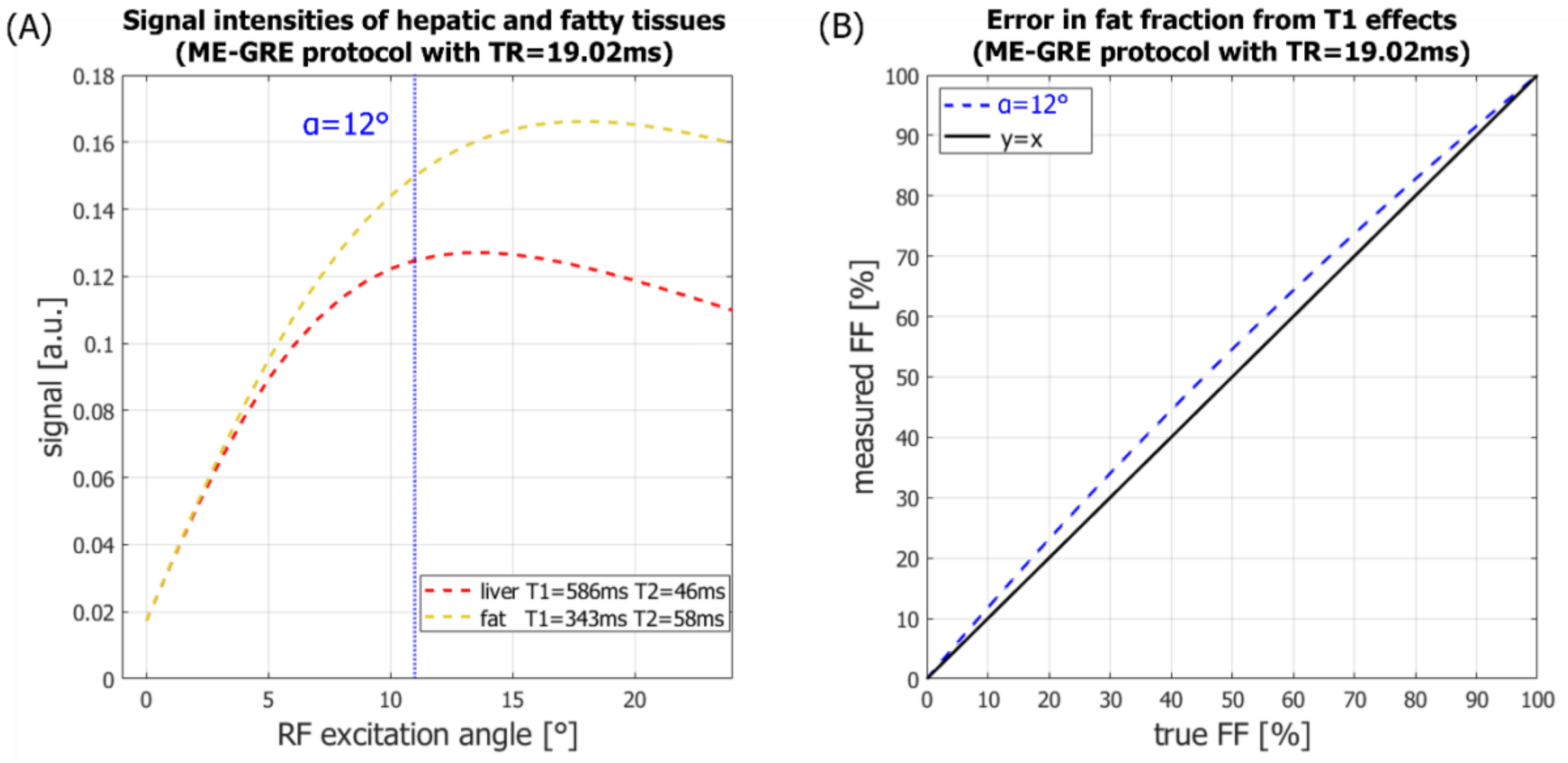

**Figure S1** **T1 bias correction curves**

Due to the relatively short T1 of fat in comparison to water-based tissues, the accelerated T1 recovery process amplifies the signal intensity of fat, resulting in apparent increases in FF. This effect is both TR- and flip-angle-dependent. To mitigate T1 bias, a common method involves low flip angle ($\alpha<5°$) acquisitions; however, this can lead to substantial SNR loss for acquisitions with numerous echoes and long TR, as employed in this work. In our study, we used a flip angle of $\alpha=12°$ for the ME-GRE scans and the method validated by Yang et al.[49] to correct for T1-induced errors in fat fraction estimations. Based on relaxometry constants reported in the literature by Bazelaire et al.[48] for normal tissues at 1.5T, a TR-adjusted correction factor of $\kappa=1.2021$ was computed and used to correct the FF estimations of the ISMRM Fat-Water Toolbox[46]. The maximal deviation from true fat fraction (B) with the proposed sequence was 4.6% for FF=47.7%.

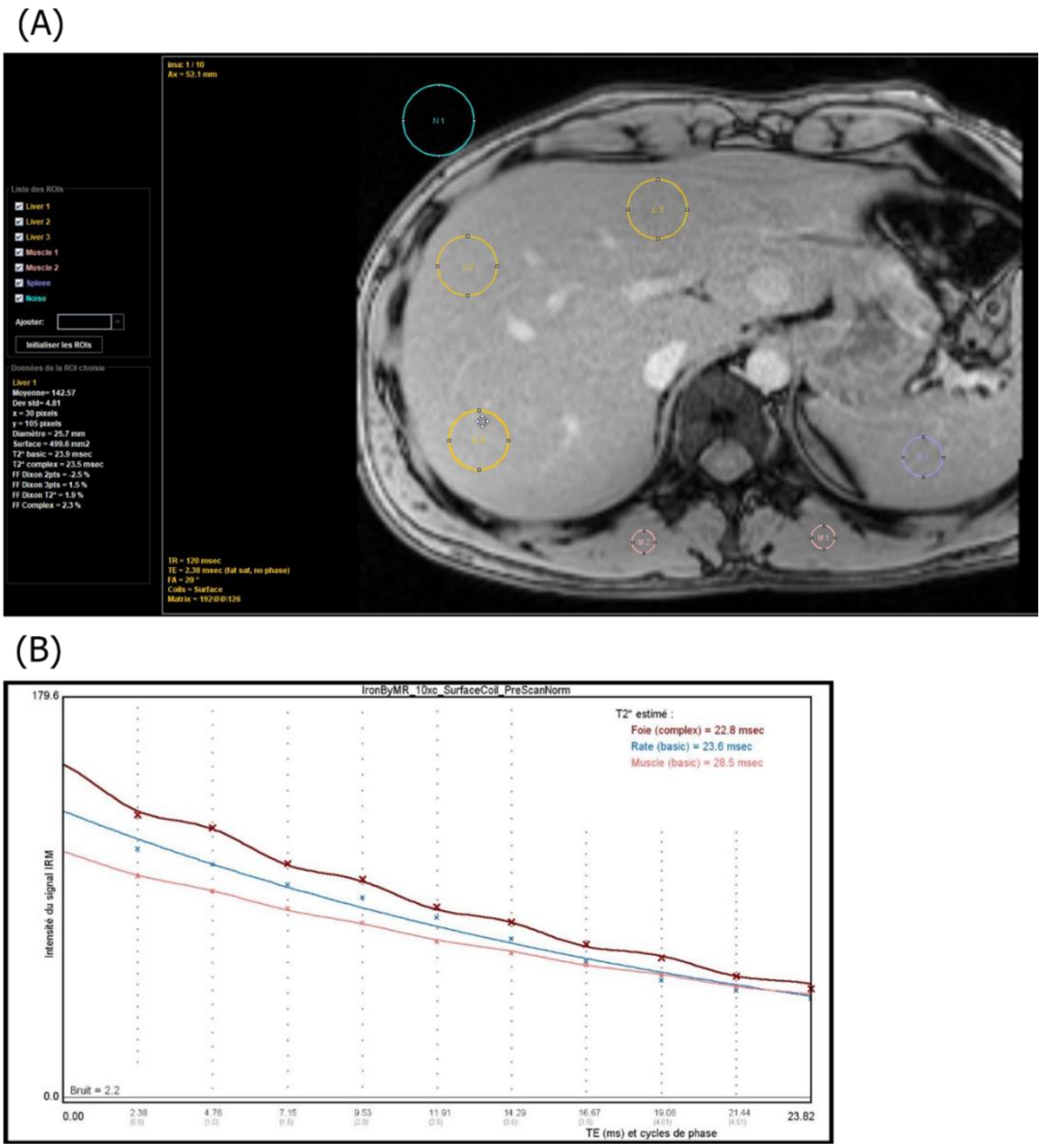

## Figure S2 Fat and iron quantification with MRQuantiF software

MRQuantiF (https://imagemed.univ-rennes1.fr/en/mrquantif/overview) is a free software developed at the University of Rennes, France. MRQuantiF is routinely used by clinicians at our institution for hepatic fat and iron quantification. The MRI protocol used in our study follows the recommended guidelines provided by the developers. After imaging with the 10-echo GRE sequence, the sources images are reconstructed at the scanner in DICOM format. The interface of the MRQuantiF processing software on a PACS workstation is shown in (A), with the three liver ROI drawn in yellow. Quantification results for the selected ROI appear in the bottom left panel. T2* decay curves are also provided for visualization (B). For fat quantification, note that no T1 bias correction is performed on the FF estimates as the MRQuantiF protocol uses a long repetition time (TR=120ms) which minimizes T1 bias despite the high flip angle of 20°.

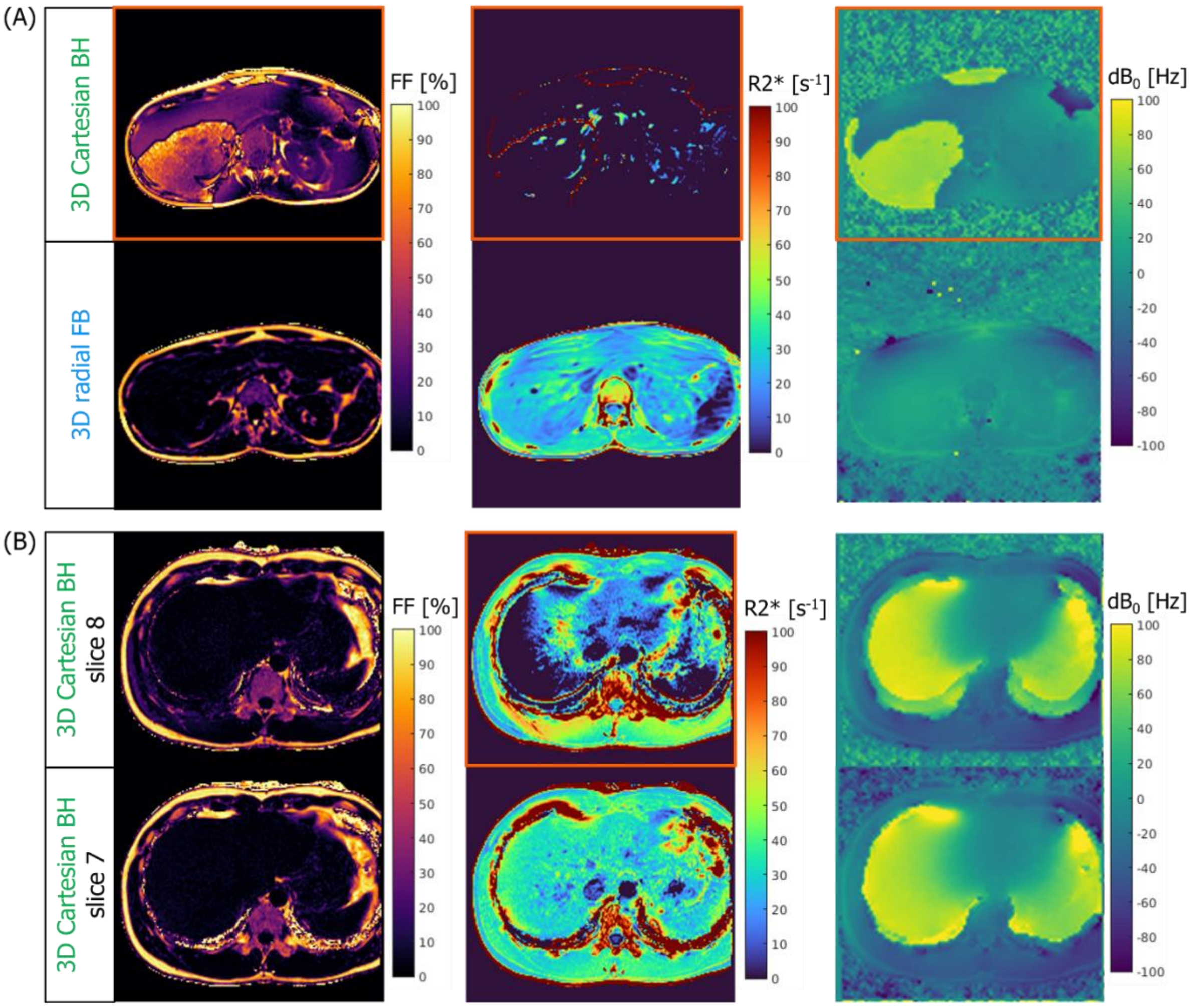

**Figure S3 Examples of corrupted FF and R2* maps obtained with a maximum-likelihood fitting routine**

Fat fraction (FF), R2* and B0 field maps obtained with a maximum-likelihood fitting routine based on graph cuts are displayed for two volunteers. Corrupted maps are highlighted by an orange frame. In panel (A), both the FF and R2* map obtained with the 3D Cartesian acquisition are corrupted, with over-estimations in FF and null R2* values found over the entire FOV. The pattern of the mis-estimations follows a patchy appearance observed in the B0 field map, indicating the failure to estimate fat and R2* is the result of a mis-estimation of the B0 field map. For reference, the corresponding (non-corrupted) maps obtained with the 3D radial FB acquisition are displayed in the second row of panel (A). In panel (B), one slice (slice 8, first row) of the R2* map obtained from the Cartesian BH acquisition showed large deviations from the values estimated in the adjacent slice (slice 7, second row), notably with null values found in the parenchyma. In this case, the FF map was not corrupted and the $B_0$ field map matched with that of the adjacent slice. In this study, all slices for which at least one map was identified as corrupted were discarded from the quantitative analyses reported.

| | 3D radial FB | | | | | | 3D Cartesian BH | | | | | | 2D MRQuantiF | | | | | |
|---|---|---|---|---|---|---|---|---|---|---|---|---|---|---|---|---|---|---|
| | ROI 1 | | ROI 2 | | ROI 3 | | ROI 1 | | ROI 2 | | ROI 3 | | ROI 1 | | ROI 2 | | ROI 3 | |
| | mean ± std | n | mean ± std | n | mean ± std | n | mean ± std | n | mean ± std | n | mean ± std | n | mean | n | mean | n | mean | n |
| **V1** | 2.9±0.6 | 4660 | 2.2±0.5 | 4024 | 2.2±0.6 | 4078 | 3.2±0.6 | 1979 | 3.0±0.6 | 1986 | 2.6±0.6 | 1931 | 2.3 | 142 | 3.0 | 129 | 1.8 | 110 |
| **V2** | 1.8±0.2 | 4248 | 2.0±0.2 | 3652 | 2.0±0.3 | 2305 | 3.5±0.9 | 701 | 3.2±0.9 | 740 | 2.0±0.5 | 555 | 2.0 | 130 | 2.0 | 114 | 2.0 | 94 |
| **V3** | 2.5±1.0 | 4220 | 2.0±0.4 | 5617 | 1.4±0.3 | 3387 | 2.7±0.8 | 2104 | 2.9±0.5 | 2815 | 2.1±0.3 | 1965 | 2.3 | 127 | 2.0 | 94 | 1.0 | 96 |
| **V4** | 8.2±0.6 | 3867 | 8.1±1.2 | 4214 | 7.4±0.6 | 3909 | 10.6±0.9 | 1996 | 9.8±1.1 | 1734 | 6.4±1.0 | 2078 | 8.9 | 119 | 8.3 | 109 | 7.1 | 96 |
| **V5** | 2.6±0.7 | 3731 | 3.4±0.4 | 5337 | 1.7±1.3 | 4066 | 3.7±0.6 | 1600 | 4.5±0.9 | 2108 | 3.9±1.0 | 2155 | 2.8 | 128 | 3.5 | 145 | 1.7 | 155 |
| **V6** | 4.6±1.1 | 4679 | 3.4±0.9 | 4881 | 4.6±0.8 | 3781 | 4.3±0.9 | 2269 | 4.2±1.0 | 2007 | 4.2±1.1 | 1091 | 3.6 | 146 | 3.3 | 129 | 3.8 | 123 |
| **V7** | 2.7±1.0 | 6097 | 1.8±0.8 | 5037 | 2.9±0.9 | 6033 | 2.7±0.9 | 2931 | 2.8±1.1 | 2779 | 3.1±0.5 | 2714 | 2.0 | 130 | 1.7 | 145 | 0.0 | 156 |
| **V8** | 1.1±0.4 | 5574 | 1.6±0.3 | 4910 | 2.1±0.2 | 4832 | 2.6±0.4 | 2195 | 2.2±0.7 | 1754 | 2.7±0.7 | 1317 | 1.1 | 147 | 1.4 | 133 | 1.2 | 139 |
| **V9** | 4.1±1.3 | 4320 | 2.5±1.0 | 4184 | 1.1±0.4 | 4333 | 3.1±1.3 | 2054 | 2.0±0.8 | 1939 | 2.6±1.7 | 2175 | 2.2 | 137 | 3.0 | 129 | 1.9 | 130 |
| **V10** | 4.7±1.0 | 5593 | 4.1±0.9 | 4506 | 2.0±0.4 | 6342 | 5.4±0.4 | 3146 | 5.5±0.6 | 2570 | 3.7±0.9 | 3011 | 4.6 | 125 | 4.2 | 109 | 2.7 | 106 |

**Table S4 Table of parametric FF measurements**

For each measurement technique (3D radial FB, 3D Cartesian BH, 2D MRQuantiF) the mean (and standard deviation, when provided) fat fraction and the number of voxels per region of interest over the whole measurement volume (respectively 20,8, and 1 slice(s) for 3D radial FB, 3D Cartesian BH and 2D MRQuantiF) are reported.